\documentclass{article}
\usepackage[dvips]{graphicx}
\usepackage{amssymb,amsfonts,amsmath}

\def\tr{\mbox{tr\hskip 1pt}}

\begin{document}

\title{Phenomenological modeling of DNA overstretching}

\author{Ray W. Ogden\\
Department of Mathematics,University of Glasgow, \\ Glasgow G12 8QW, Scotland, UK\\
\texttt{e-mail:rwo@maths.gla.ac.uk}\\ Giuseppe Saccomandi \\
Dipartimento di Ingegneria Industriale,\\
Universit\`{a} degli Studi di Perugia, 06125 Perugia, Italy \\
\texttt{giuseppe.saccomandi@mec.dii.unipg.it} \\ Ivonne Sgura \\
Dipartimento di Matematica,\\
Universit\`{a} degli Studi di Lecce, 73100 Lecce, Italy\\ \texttt{e-mail:ivonne.sgura@unile.it}}

\date{\today}

\maketitle

\begin{abstract}
A phenomenological model based on the three-dimensional theory of
nonlinear elasticity is developed to describe the phenomenon of
overstretching in the force-extension curve for dsDNA. By using
the concept of a material with multiple reference configurations a
single formula is obtained to fit the force-extension curve.
\end{abstract}
{\bf Abbreviations:} dsDNA, double stranded DNA; ssDNA, single stranded DNA; WLC, Worm like chain.

\section{Introduction}

A typical force-extension curve for dsDNA exhibits three \textit{portions} \cite{Bouchiat,SCroquette,Marko95}. During
the first portion there is an entropic stretching regime (usually modeled by the worm-like chain), followed by a force
plateau in the region of $65$ picoNewtons, while in the last portion there is a sharp transition from the usual B-form
to a new overstretched form, usually designated S-DNA. The structure of S-DNA remains the subject of debate but it
should not be confused with ssDNA \cite{Williams,Sdna}. The biological function of the overstretching DNA transition is
complementary to thermal or pH induced denaturation and for this reason there is considerable attention focused on
modeling this phenomenon \cite{Metzler}, but there appears to be no general agreement about the models that have been
proposed in the literature.

In \cite{Rouzina1} a model based on a force-induced melting of the
DNA double helix was proposed. This implies that S-DNA is made up
of a mixture of large islands of separated ssDNA and remnant
base-paired B-DNA, and molecular extension is a weighted average
of its extension in the two possible states. A two-state worm-like
chain has been also proposed in \cite{Ashan}. In several papers
the idea that the overstretching behavior of DNA may be modeled
by a sort of \textit{mixture} theory has been applied in the study
of B-DNA to S-DNA transition as a function of solution conditions,
including variations in temperature, pH and ionic strength (see,
for example, \cite{Rouzina2}). In \cite{Cocco}, by using a
thermodynamical model for tension-melted dsDNA it is argued that
the overstretching transition cannot be explained in terms of
\textit{conversion} of double helix to noninteracting
polynucleotide strands. This is because two parallel
noninteracting ssDNAs cannot explain quantitatively the mechanical
properties of S-DNA. This is argued directly from an examination
of the experimental data by the authors \cite{Cocco}. The Rouzina
and Bloomfield model \cite{Rouzina1, Rouzina2} is therefore
criticized because in the B--ss scenario the overstretched state
should be associated with a constant force between the B-DNA
and ssDNA lower than that observed.

The aim of the present note is to present a new framework for
describing the overstretching phenomena. Our model is
developed using a non-standard version of the phenomenological
theory of nonlinear elasticity where the stress is
determined as a function of the deformation gradient calculated
with respect to a \textit{varying reference
configuration} in a such a way that it is possible to introduce
micro-mechanical considerations. This idea was
introduced originally by Eckart \cite{Eckart} and then developed
more recently by Rajagopal and Wineman \cite{Raja,Wineraja} in
order to formulate constitutive equations
for materials that undergo deformations induced by
microstructural changes. Recently De Tommasi et al. \cite{Rheology}
have proposed a micro-mechanical interpretation of
this theory that may be quite useful in the study of the
overstretching phenomenon.

The general form of the constitutive equation for nonlinear
elasticity is expressed in terms of a strain-energy function. In
the standard theory it is assumed implicitly that the material
response is due to a molecular mechanism that does not change
during the deformation process under consideration. In
single-molecule experiments on DNA, this assumption may be
considered valid only on the first portion of the force-extension
curve. At a certain moment the hydrogen bonds between strands
start to break and there is a fundamental change in the molecular
mechanism responsible of the overall material response. In the
case of DNA these microstructural changes are driven by several
factors: stretching, salinity, temperature, etc. From a
micro-mechanical point of view it is possible to look at the
nucleotides as particles that are connected by two different types
of chains. A fraction of the chains is elastic and endows the DNA
molecule with nonzero stiffness. The complementary chains are
breakable and are responsible for the alteration of the molecule.
The stress in each breakable chain is zero until a certain
activation threshold is reached and after a limiting value of the
strain is overcome. We assume that a continuous process of
microstructural conversion occurs after the deformation increases
beyond a threshold value. In this initial model we shall neglect
any factors such as salinity, temperature or ionic strength that
are not strictly mechanical. We emphasize that instead of
considering the \textit{conversion} of the double helix to two
noninteracting polynucleotide strands, we are considering here a
process of conversion related to the rupture of stress-bearing
bonds. Upon rupture of the bonds a new microstructural arrangement
forms with a new unstressed reference configuration. More details
of this constitutive model may be found in \cite{Rheology}.

Let us consider a deformation
$\mathbf{x}=\mathbf{x}(\mathbf{X},t)$, where $\mathbf{x}$ is the
current position of a particle located at $\mathbf{X}$ in the
undeformed configuration at time $t=0$. The deformation gradient
is given by $\mathbf{F}(\mathbf{X},t)=\partial \mathbf{x/}\partial
\mathbf{X}$ and the left Cauchy-Green tensor by
$\mathbf{B}=\mathbf{F}\mathbf{F}^{T}$.  We assume that there is a
range of deformation for which the material behaves like an
incompressible, isotropic elastic material, i.e. the Cauchy stress
$\mathbf{T}=-p\mathbf{I}+\mathbf{T}^{(\text{E,1})}$, where
$-p\mathbf{I}$ is the indeterminate part of the stress due to the
constraint of incompressibility ($\det \mathbf{F}=1$) and the
extra stress takes the form
\begin{equation}
\mathbf{T}^{(\text{E,1})}=2W_{1}^{(1)}\mathbf{B}-2W_{2}^{(1)}\mathbf{B}^{-1}.
\label{p0}
\end{equation}
The strain-energy function $W^{(1)}=W^{(1)}(I_{1},I_{2})$ is a function of the principal invariants $I_{1}=\tr(
\mathbf{B})$ and $I_{2}=\tr( \mathbf{B}^{-1})$, where $W_{i}^{(1)}=\partial W^{(1)}/\partial I_{i},\,i=1,2$. An
activation criterion is needed to determine when the microstructural change begins. This is provided by introducing a
scalar deformation state parameter $s$. Here, we suppose that $s=s(I_{1},I_{2})$ depends on the deformation through
$I_1$ and $I_2$ for consistency with the requirement of isotropy, although more general forms of $s$ may easily be
adopted. For $s < s_{a}$, the threshold value of $s$, no conversion has yet occurred, i.e. all the material is in its
original form and the stress is given by \eqref{p0}. On the other hand, for a value of the state parameter
$\widehat{s}$ beyond $s_{a}$ microstructural changes have occurred and the reference configuration has changed. This
implies that the stress is now a function of the relative deformation gradient for the material formed at state
$\widehat{s}$ given by $\widehat{\mathbf{F}} =\partial \mathbf{x}/\partial \widehat{\mathbf{x}}$, where
$\widehat{\mathbf{x}}$ is the position of the particle in the configuration corresponding to deformation state
$\widehat{s}$. In Figure 1 the original reference configuration, the configuration at $\widehat{s}$ and the current configuration are depicted. The associated Cauchy-Green tensor is given by $\widehat{\mathbf{B}}=\widehat{\mathbf{F}}
\widehat{\mathbf{F}}^{T}$.

\begin{figure}
\begin{center}
\includegraphics*[width=7cm, height=6cm]{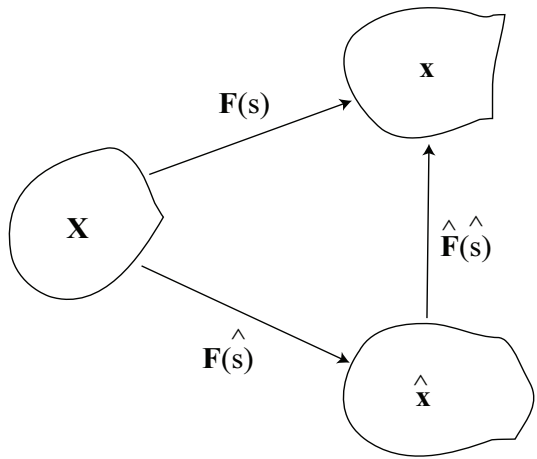}
\end{center}
\caption{Schematic of a material with an evolving reference
configuration}
\end{figure}

We shall assume that the new material formed at the state
$\widehat{s}$ is still elastic, isotropic and incompressible such
that the extra Cauchy stress at state $s$ in this new
configuration formed at the deformation state $\widehat{s}$ is
given by
\begin{equation}
\mathbf{T}^{(\text{E,2})}=2W_{1}^{(2)}\widehat{\mathbf{B}}-2W_{2}^{(2)}
\widehat{\mathbf{B}}^{-1}.  \label{p00}
\end{equation}
Here $W^{(2)}=W^{(2)}(\widehat{I}_{1},\widehat{I}_{2})$ is the
strain-energy function of the newly formed material, relative to
the reference configuration at $\widehat{s}$. Another important
simplifying assumption is that a single function $W^{(2)}$ governs
the strain energy during the continuous microstructural change.
The total current stress is taken as the superposition of the
contributions from the material remaining in its original
configuration and from all the new material formed at deformation
states $\widehat{s}\in \left[ s_{a},s\right]$, i.e.
\begin{equation}
\mathbf{T}=-p\mathbf{I}+b(s)\mathbf{T}^{(\text{E,1})}
+\int_{s_{a}}^{s}a(\widehat{s})\mathbf{T}^{(\text{E,2})}d\widehat{s}.
\label{p01}
\end{equation}
In \eqref{p01} the function $a(s)$ is a conversion rate satisfying
$a(s)=0$ when $s\leq s_{a}$ and $a(s)>0$ for $s>s_{a}$, while
$b(s)$ is the volume fraction of the material in the original
configuration remaining at state $s$, with $b(s)=1$ when $s\leq
s_{a}$ and $0\leq b(s)<1$ for $ s>s_{a}$.  Thus, to complete the
model, constitutive equations for $W^{(1)}$ and $W^{(2)}$, the
activation criterion and the conversion rate have to be
prescribed. Our model is three-dimensional and fully consistent
with the theory of continuum mechanics. To illustrate the ideas
quantitatively we begin with a prototype that is empirical and
one-dimensional, and we then show how to recast the theory in
three-dimensional form.

\section{The constitutive models}

\subsection{Data sources}

\ We consider the sets of experimental data in \cite{Williams,Sdna},
which correspond to different salt concentrations. We use measured
data $(x_i,f_i),\, i=1,\dots, m$, here corresponding to a
force-($f$ in picoNewtons)-extension ($x$ in microns $\mu$) experiment
on a single dsDNA molecule in 250\,mM [Na$^+$] buffer solution at
7.5\,pH (these data are reported in figure 3 of \cite{Williams}).

\subsection{An empirical one-dimensional model}

\ Let $x$ denote the one-dimensional extension. On the same basis
as illustrated above, for the one-dimensional force $f$ we have
\begin{equation}
f(x)=b(x)f^{(1)}+\int_{x _{a}}^{x }a(\widehat{x})f^{(2)}\left( \widehat{x }%
\right) d\widehat{x}.  \label{e1}
\end{equation}
In \eqref{e1} the various quantities $a(x), b(x), x_{a}$ have the
same meaning as before, with $x$ replacing $s$.  Since the process of
conversion is continuous we have
\begin{equation}
b(x)=1-\int_{x_{a}}^{x}a(\widehat{x})d\widehat{x},\quad x\geq x_a. \label{p8bis}
\end{equation}

The constitutive assumptions we introduce are: for $f^{(1)}$, a
logistic modification of the original one-dimensional Fung model
widely used in biomechanics \cite{Fung}, i.e.
\begin{equation}
f^{(1)}(x)=\frac{\mu _{1}}{2}\frac{\exp [\beta(x-x_{0})]}{\exp
[\beta(x-x_{0})]+\gamma },  \label{fungone}
\end{equation}
where the material constants $\mu_1$, $x_0$ and $\beta$ have dimensions
of force, length and $1$/length, respectively, and $\gamma>0$ is a
dimensionless constant; and, for $f^{(2)}$, the WLC
interpolation formula
\begin{equation}
f^{(2)}(x) =\mu _{2}\left[ \frac{1}{4}\left( 1-z\right)
^{-2}-\frac{1}{4}+z \right] .  \label{WLC}
\end{equation}
Here $\mu _{2}={kT/l_{p}}$, where $l_{p}$ is the persistence
length, $k$ is Boltzmann's constant and $T$ the temperature
(degrees Kelvin), $z={x/l_{c}}$ and $l_{c}$ is the contour length
of the molecule. We have chosen a logistic Fung model to capture the
first portion of the force-extension curve (i.e. to capture the
strain-hardening phenomenon) but without introducing a singularity
such as that in the WLC formula.  The WLC is used to model the
sharp increase in force at the end of the curve just after the
plateau. It is clear that the modeling of the plateau zone
depends on how the reference configuration evolves, and this may
be controlled by the choice of the conversion rate. Usually, in the
context of rubber mechanics, very simple models for the conversion
function are adopted (for example, quadratic or piecewise
linear functional forms). Here we use a functional form
suggested by statistical mechanics, namely a probability
distribution function computed by considering two possible states
for a chain composed of a fixed number of base pairs with a given
fixed difference in the energy between the two states. For this
purpose let
$$
g(x) = {\frac {\delta\,{c_1}\,{e^{-{c_1}\, \left( x-{
c_2} \right) }}} { [ 1+{e^{-{c_1}\, \left( x-{c_2}
\right) }}] ^{2}}},
$$
where $c_1,c_2$ and $\delta$ are constants, and define
\begin{equation}
a(x) = g(x) - g(x_a) \qquad  x \in [x_a, x_c], \label{barbi}
\end{equation}
with $a(x)=0$ otherwise. Here, we are assuming that the conversion has
been completed when $x$ reaches the value $x_c$, and this imposes the
continuity requirement $a(x_c)=0$, which leads to $c_2 = (x_a + x_c)/2$.
A plot of the function $a(x)$ is shown in Figure 2.

We denote by $C$ the total fraction of the material that can undergo conversion.
Then,
\begin{equation}
C=\int_{x_{a}}^{x_c}a(\widehat{x })d\widehat{x}. \label{p7bis}
\end{equation}

From the definition of $C$ in \eqref{p7bis}, we calculate
$$
\delta/C =  {\frac {  [ 1+{e^{- {c_1}\, \left( x_c -{
c_2} \right) }} ] [ 1+{e^{-{c_1 }\, \left( x_a-{
c_2} \right) }} ] } {\, {e^{-{c_1}\, \left( x_a-{
c_2} \right) }}- {e^{-{c_1}\, \left( x_c-{c_2} \right) }}
}}- g(x_a)(x_c -x_a).
$$
The constitutive parameters to be found in this empirical model
are $\mu _{1},\beta,\gamma,\mu _{2}$ and $l_c$. Moreover, we have
to fix the activation criterion and therefore we also need values
for $x_{a}$, $x_c$, $c_1$ and $C$. At this stage the only \emph{a
priori} information about these parameters is that $C\in [0,1]$.
The strategy for fitting that we use to deal with the original
force-extension data $(x_i,f_i), i=1, \dots, m$, is explained in
the Appendix. Since, in principle, several parameters have to be
identified in our model, their numerical approximation could pose
severe problems (see, e.g., \cite{compmech}). For this reason we
devise a strategy that accounts for the physical interpretation of
some of these parameters. A set of parameters identified by the
fitting results is given by
\begin{equation}
\left.
\begin{array}{lll}
x_a = 20, &     x_c = 28.754, &  C = 0.50367, \\[0.1cm]
\mu_1 = 64.977, &  \beta= 2.7537,&    \gamma = 0.019288,\\[0.1cm]
c_1 = 0.059145, &    \mu_2 = 25.87, & l_c= 32.022,
\end{array}\right\}\label{fit1}
\end{equation}
with residual $\text{res}^2=21.93$.

\begin{figure}
\begin{center}
\includegraphics*[width=10cm, height=7cm]{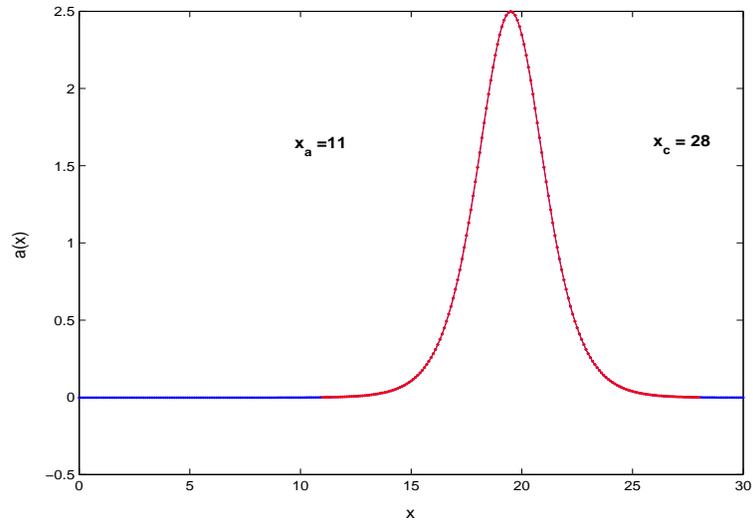}
\end{center}
\caption{Activation criterion with $a(x)$ plotted against $x$, illustrated for $x_a=11$ and $x_c =28$ with $c_1 =1$,
based on the set of data in Wenner et al. \cite{Williams} for 250\,mM}
\end{figure}

In Figure 3 the prediction of the model obtained by using these
parameters is shown. The results are quite good, but we believe
that better insight might be gained from the three-dimensional
model.

\begin{figure}
\begin{center}
\includegraphics*[width=9cm, height=7cm]{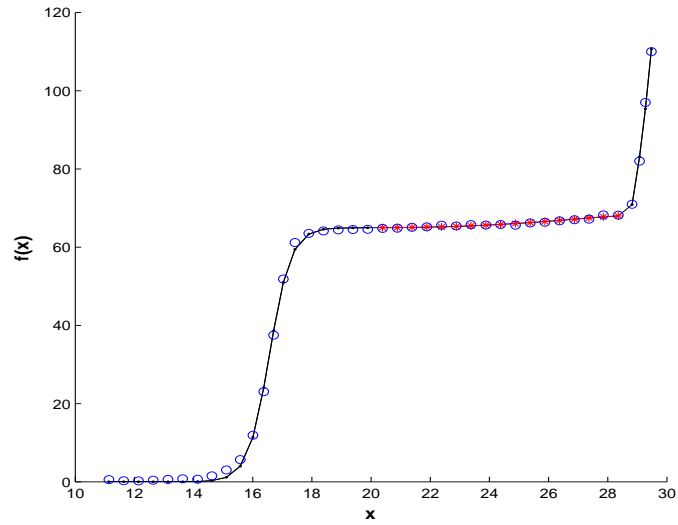}
\end{center}
\caption{Plot of $f(x)$ vs. $x$. Circles are from the data in Wenner et al.
\cite{Williams} for 250\,mM.  The curve is
obtained from the fitting procedure, which yields the parameter values given in \eqref{fit1}
with squared residual res$^2 = 21.93$.  The red stars indicate the range of values for which the conversion
is active}
\end{figure}

\subsection{Three-dimensional models}

\ In three dimensions the single molecule force-extension experiment
is idealized as a simple tension test, for which the deformation is given by
\begin{equation}
x=\frac{1}{\sqrt{\lambda }}X,\quad y=\frac{1}{\sqrt{\lambda
}}Y,\quad z=\lambda Z,  \label{p1}
\end{equation}
where $\lambda $ is the stretch in the axial (i.e. $z$) direction. The current deformation gradient is given in matrix
form by $\mathbf{F}(\lambda)=\text{diag}( 1/\sqrt{ \lambda },1/\sqrt{\lambda },\lambda )$, and the corresponding
Cauchy-Green deformation matrix is $\mathbf{B}(\lambda)=\text{diag}( 1/\lambda ,1/\lambda ,\lambda ^{2})$. Hence,
\begin{equation}
I_{1}=\lambda ^{2}+2\lambda ^{-1},\quad I_{2}=\lambda
^{-2}+2\lambda . \label{I1}
\end{equation}

Since this is a one-parameter deformation, it is possible to
establish that there is a one-to-one correspondence between the
activation parameter $s$ and the stretch, and we write
$s=s(\lambda)$.  For this reason we use the terminology
\emph{activation stretch}, which we denote by $\lambda _{a}$,
instead of a generic activation parameter $s_{a}$ (then,
$s_{a}=s(\lambda _{a})=\lambda _{a}$). The deformation gradient at
state $\widehat{\lambda }$ is therefore denoted by
$\widehat{\mathbf{F}}(\widehat{\lambda })=\mathbf{F}(\lambda
)\mathbf{F} ^{-1}(\widehat{\lambda })$ and we therefore compute
\begin{eqnarray}
&& \widehat{\mathbf{F}}(\lambda )=\text{diag}\left(
\sqrt{\widehat{\lambda } /\lambda },\sqrt{\widehat{\lambda
}/\lambda },\lambda /\widehat{\lambda }
\right), \nonumber\\
&& \widehat{\mathbf{B}}(\lambda )=\text{diag}\left( \widehat{
\lambda }/\lambda ,\widehat{\lambda }/\lambda ,\lambda
^{2}/\widehat{\lambda }^{2}\right) .  \label{p2}
\end{eqnarray}
It follows that, for example, $\widehat{I}_{1}=\lambda ^{2}/
\widehat{\lambda }^{2}+2\widehat{\lambda }/\lambda$. If we
consider the class of elastic materials referred to as generalized
neo-Hookean materials, with $W=W(I_{1})$, then from \eqref{p0} we
obtain the principal components of the Cauchy stress tensor in the
form
\begin{equation}
t_{i}=2\lambda _{i}^{2}W_{1}-p,\quad i=1,2,3.  \label{p3}
\end{equation}
The requirement that the lateral surfaces of the specimen
undergoing the simple extension are traction free,
$t_{1}=t_{2}=0$, yields
\begin{equation}
p=2\lambda ^{-1}W_{1}.  \label{p4}
\end{equation}
Generalizing these results to the case \eqref{p01} the tensile
force per unit deformed cross-sectional area necessary to
achieve the stretch is given by the Cauchy stress component
\begin{eqnarray}
t_{3}(\lambda )&=&2b(\lambda )\left( \lambda ^{2}-\lambda
^{-1}\right)
W_{1}^{(1)} \nonumber \\
&+&2\int_{\lambda _{a}}^{\lambda }a(\widehat{\lambda })\left(
\frac{ \lambda ^{2}}{\widehat{\lambda
}^{2}}-\frac{\widehat{\lambda }}{\lambda } \right)
W_{1}^{(2)}\left( \widehat{\lambda }\right) d\widehat{\lambda },
\label{p5}
\end{eqnarray}
where
\begin{equation}
b(\lambda
)=1-\int_{\lambda_{a}}^{\lambda}a(\widehat{\lambda})d\widehat{\lambda}.
\label{p16}
\end{equation}
The corresponding force per unit undeformed area of cross-section is
$\lambda^{-1}t_3(\lambda)$.

At this point it is necessary to complement \eqref{p5} with the
constitutive equations. We need a constitutive equation for the
strain-energy function of the material before the conversion
starts, i.e. $W^{(1)}$, to model the first portion of the
force-extension curve. Then, we also need a constitutive
equation for the function $W^{(2)}$ that governs the mechanical
behavior of the newly formed material. This choice is important
for modeling the ``last'' portion of the force-extension curve.
The overstretching plateau, as already pointed out, is modeled by
the choice of the conversion function $a(s)$. For the strain
energy $W^{(1)}$ in the first regime we consider a
modification of the strain-energy function, here denoted $W^F$,
proposed by Fung for modeling biological tissues.  This is given by
\begin{equation}
W_{1}^{F}=\frac{\mu }{2}\exp [\beta(I_{1}-3)].  \label{fungW1}
\end{equation}
As for the 1D case, we need to modify this relationship because a
saturation phenomenon has to be taken into account. The mechanical
behavior characterizing the strain stiffening of the DNA molecule
in the first portion of the force-extension curve cannot influence
what happens in the plateau zone. For this reason we consider a
logistic modification of the (three-dimensional) Fung model
\eqref{fungW1} analogous to that used for 1D.  This is given by
\begin{equation}
W_{1}^{(1)}=\frac{\mu _{1}}{2}\frac{\exp [\beta(I_{1}-3)]}{\exp
[\beta(I_{1}-3)]+\gamma },  \label{fungmodW1}
\end{equation}
so that
\begin{equation}
W^{(1)}=\frac{\mu _{1}}{2\beta}\ln \left( \exp (\beta (I_{1}-3))+\gamma \right), \label{fungmodW}
\end{equation}
which reduces to the neo-Hookean material $W=\mu_1 (I_{1}-3)/2$ when
$\gamma =0$. For the strain-energy function
$W^{(2)}$ in the second portion of the deformation range we consider
the phenomenological model first proposed by Gent
\cite{Gent} and given by
\begin{equation}
W^{(2)}(\widehat I_1) =-\frac{\mu_2 }{2}J_m\ln \left(
1-\frac{\widehat I_1-3 }{J_m}\right) ,\quad \widehat I_1<J_m+3,
\label{g1}
\end{equation}
where $\mu_2 $ is the shear modulus for infinitesimal deformations
and $J_m\,(>0)$ is the limiting value of $\widehat I_1-3$
associated with limiting chain extensibility. In the limit as the
chain extensibility parameter tends to infinity $\left(
J_m\rightarrow \infty \right)$, \eqref{g1}  also reduces to the
classical neo-Hookean model. The model \eqref{g1} has been
discussed in detail by Horgan \& Saccomandi \cite{HS2} and it can
be connected with the so-called Freely Jointed Chain (FJC) model.
In this case the response function is given by
\begin{equation}
W_1^{(2)} =\frac{\mu_2 }{2} \frac{J_{m}}{J_{m}-(\widehat I_{1}-3)}, \label{gent}
\end{equation}
so that the stress has a singularity as $\widehat I_{1}\rightarrow J_{m}+3$.

The model we have proposed contains
several constitutive parameters that have to be found by using
a fitting procedure. The parameters needed to fix the
strain-energy functions are $\mu _{1},\beta,\gamma $ and
$\mu _{2},J_{m}$. Moreover, we use the same activation
criterion as was used for the 1D model in \eqref{barbi}.
Hence, to fix the activation criterion we need to identify the
interval with ends $x_a \rightarrow \lambda _{a}$ and $x_c \rightarrow \lambda_{c}$
and the parameter $c_1$. Note that
this criterion may easily be reformulated in a way compatible with 3D
elasticity in terms of the invariant $I_1$. Equation
\eqref{p5} provides a formula for the Cauchy stress, but it is the nominal stress
$\lambda^{-1}t_3(\lambda)$ (force per unit reference cross-sectional area)
that is needed for the data fitting.  We therefore transform the data set $(x_i,f_i)$
into the data set $(\lambda_i,f_i)$, where $\lambda_i = 1 + x_i / l_c$,
with the contour length $l_c$ identified in the 1D case.
To match the dimensions of the force $f$ in the data the stress
$\lambda^{-1}t_3(\lambda)$ has to be multiplied by the reference cross-sectional area,
which is unknown.  However, this is just a multiplicative factor that is accounted for
by incorporating it into the constants $\mu_1$ and $\mu_2$, which then have dimensions
of force as in the 1D situation.

The parameters obtained by the strategy explained in the Appendix are
\begin{equation}
\left.
\begin{array}{lll}
\lambda_a = 1.3833, &     \lambda_c = 1.895, &   C = 0.71658, \\[0.1cm]
\mu_1 = 60,&    \beta = 43.772, &    \gamma = 1.193\text{e}5,\\[0.1cm]
c_1 = 9.6293,&    \mu_2 = 5,&  J_m= 1.8819,
\end{array}\right\}\label{fit2}
\end{equation}
with squared residual $\text{res}^2 = 57.68$. The result of this fitting
is shown in Figure 4.

\begin{figure}[h]
\begin{center}
\includegraphics*[width=11cm, height=9cm]{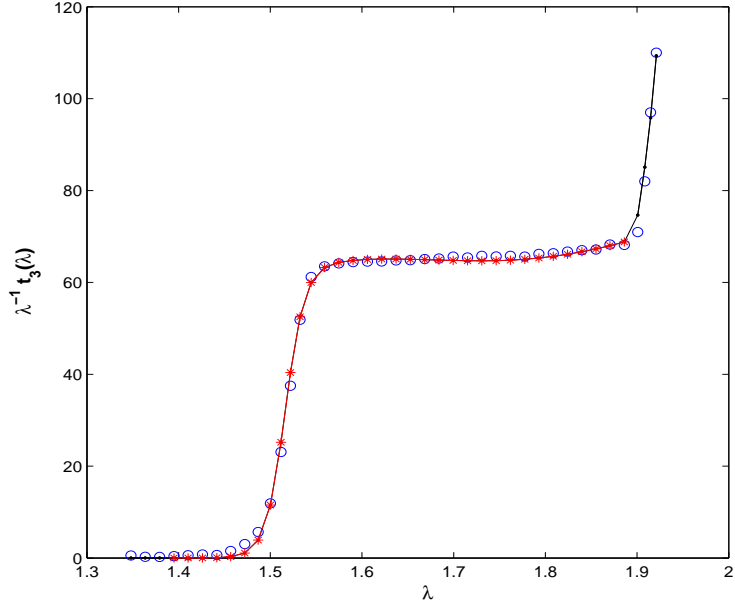}
\end{center}
\caption{Plot of $\lambda^{-1}t_3(\lambda)$ vs. $\lambda$.
Circles are from the data in Wenner et al. \cite{Williams} for 250\,mM.  The curve is
obtained from the fitting procedure, which yields the parameter values given in \eqref{fit2}
with squared residual res$^2 = 57.68$.  The red stars indicate the range of values for which
the conversion is active.  Note that the dimensions of the stress have been converted into `force'
by multiplication by the unknown reference cross-sectional area by incorporating this into
the parameters $\mu_1$ and $\mu_2$}
\end{figure}

The model proposed herein gives good results in fitting the data,
and because it has been formulated within a very general framework
it may easily be extended to take into account several variables
of biological interest. The model is interesting not only because
it is comprises a single formula describing the complete
force-extension curve, but also from a conceptual point of view.
Indeed, as has been argued by Cocco et al. \cite{Cocco}, S-DNA
cannot be described as a simple sort of mixture between the dsDNA
and ssDNA. The relationship is more complex and it is clarified by
the existence of multiple references configurations. We point out
that because DNA overwinds when it is stretched, we need a three
dimensional model to obtain a complete and realistic picture of the
single molecule experiments and our model is just a rigorous
version of the toy model proposed by Gore et al. in \cite{Gore}.

\section{Appendix}

The strategy for fitting the theoretical model to the experimental
data is based on a nonlinear least squares (LS) approximation as
follows. As a first step, we fix \emph{a priori} some parameters
from simple biological considerations and we solve the
optimization problem for the remaining parameters
in order to identify a first optimal subset, $\mathbf{p}^*$ say.
In the successive steps, the strategy consists of implementing the
LS algorithm by starting from this solution and then moving in a
descent direction by including each time a new free parameter from
amongst those that were fixed.  The solution found at each step is
then used as an initial guess for solving the next LS problem in
which a further parameter has to be identified. Only $x_c $ (for
the 1D model) and $\lambda_c$ (for the 3D model) are always fixed.
This assumption implies that in the final part of the experimental
curve for $x > x_c$ (or $\lambda > \lambda_c$) the material is all
converted to its new form.

All the computations are performed in Matlab with the
\emph{lsqcurvefit} routine (see \cite{matlab}) for solving
nonlinear least squares problems. We allow the algorithm to
perform a maximum of $3000$ iterations and stop with stringent
tolerances on the errors (tol\,$=$\,1e$-$12).

For the 1D model, in the first step we use the optimization
procedure to identify the parameters $\mathbf{p}=[\mu_1, \beta,
\gamma]^T$, while the others are fixed by considering the
following physical features:\\
-- the total contour length $l_c$ is chosen to be slightly larger
than the last datum value for the extension since its value locates the asymptote
of the WLC;\\
-- $x_0=x_a$ since the value of $x_0$ in the logistic function
\eqref{fungone} identifies the point where the largest growth
occurs, and this corresponds to the meaning of $x_a$ in the
activation criterion;\\
-- since  $\mu_1$ corresponds to the horizontal asymptote of the
logistic function, its starting guess is set to almost
the force value of the plateau in the data; thus, we set $\mu_1 =69$;\\
-- we set $x_a =20$ so that up to the beginning of the plateau the
material is all in its original form. Moreover, we set $C=0.5$,
requiring by this assumption a conversion of $50\%$.

The optimal parameter set identified in this first step, is $ \mathbf{p}^*=[\mu_1^*, \ \beta^*, \ \gamma^*]^T=[69.0883,
\ 2.907, \ 0.0153]^T$. Hence, by using this first approximation, we define a new sequence of optimization problems,
where the fixed parameters are considered free in the (arbitrary) sequence $[C, \lambda_a,  l_c,  c_1,  \mu_2]$. The
optimal final result is reported in the text and in Figure 3.

The same fitting strategy used for fitting the data with the
1D empirical model \eqref{e1} is used for the 3D model \eqref{p5}. For the
activation criterion we fix $\lambda_c =1.895$.  Moreover,
we set $C = 0.8$. For the Gent material we fix $\mu_2$ equal to almost the force corresponding to the plateau, and
the parameter $J_m$, accounting for the asymptote location, such
that $J_m$ is almost the last numerical value available for the stretch data.
At the first stage of the fitting the free parameters are again those
of the Fung model and if $\mathbf{p^*}$  is the set identified in this step, the (arbitrary) sequence
in which the other parameters are considered as free is $[C, J_m, \lambda_a, c_1, \mu_2]$. In
this way a better (lower residual) optimal solution is found and
the result is reported in the text and in Figure 4.

\section{Acknowledgments}

This work was partially supported by the Italian MIUR under a project PRIN 2004 $M^2XD^2$. We are grateful to Maria
Barbi for suggesting the activation criterion and to Kumbakonam Rajagopal for interesting discussions on generalized
continuum mechanics. We wish to thank Mark Willams and Victor Bloomfield for furnishing us with the values of experimental data in \cite{Williams}.

\end{document}